\documentclass[twocolumn,twoside]{IEEEtran}

\usepackage{graphicx}
\usepackage{cite}
\usepackage{enumerate}

\usepackage{setspace}

\usepackage[dvips]{epsfig}
\usepackage{epstopdf}

\usepackage{algorithm}
\usepackage{algorithmicx}
\usepackage{algpseudocode}

\usepackage{multirow}

\usepackage{subcaption}
\usepackage{wrapfig}

\usepackage{float}

\usepackage{bm}

\usepackage{amsmath}

\makeatletter
\usepackage{amsthm}
\usepackage{amsbsy}

\usepackage{array}
\usepackage{fancyhdr}

\usepackage{color}

        \def\@oddhead{\hbox{\scriptsize ACCEPTED FOR PUBLICATION IN IEEE WIRELESS COMMUNICATIONS LETTERS}\scriptsize\rightmark \hfil \thepage}
        \def\@evenhead{\scriptsize\thepage \hfil \leftmark\hbox{\scriptsize ACCEPTED FOR PUBLICATION IN IEEE WIRELESS COMMUNICATIONS LETTERS}}

\begin{document}
\title{On the Co-existence of TD-LTE and Radar over 3.5 GHz Band: An Experimental Study}
\author{Jeffrey H. Reed, Andrew W. Clegg, Aditya V. Padaki, Taeyoung Yang, Randall Nealy, Carl Dietrich, \\Christopher R. Anderson, and D. Michael Mearns \vspace{-24pt} \thanks{This paper has been accepted for publication in IEEE Wireless Communications Letters. This work was supported by Google, Federated Wireless, NSF (ECCS-1247928), and Wireless@VT Affiliates. 

J. H. Reed, A. V. Padaki, T. Yang, R. Nealy, and C. Dietrich are with Wireless@VT, Bradley Department of Electrical and Computer Engineering, Virginia Tech, Blacksburg, VA. A. W. Clegg is with Google, Reston, VA. This work was carried out when A. V. Padaki was with Google and T. Yang was with Virginia Tech. C. R. Anderson is with the US Naval Academy, Annapolis, MD, and D. M. Mearns is with the Naval Surface Warfare Center, Dahlgren, VA. Corresponding author is A. V. Padaki.}

}

\maketitle 

\begin{abstract}

This paper presents a pioneering study based on a series of experiments on the operation of commercial Time-Division Long-Term Evolution (TD-LTE) systems in the presence of pulsed interfering signals in the 3550-3650 MHz band. TD-LTE operations were carried out in channels overlapping and adjacent to the high power SPN-43 radar with various frequency offsets between the two systems to evaluate the susceptibility of LTE to a high power interfering signal. Our results demonstrate that LTE communication using low antenna heights was not adversely affected by the pulsed interfering signal operating on adjacent frequencies irrespective of the distance of interfering transmitter. Performance was degraded only for very close distances (1-2 km) of overlapping frequencies of interfering transmitter.\vspace{-3pt}

\end{abstract}


\vspace{-6pt}
\section{Introduction}

In 2012, the President's Council of Advisors on Science and Technology (PCAST)\cite{PCAST12} recommended sharing of 1,000 MHz of radio frequency spectrum that is currently used only for U.S. Government purposes, as a means of stimulating economic growth and ensuring U.S. leadership in spectrum sharing technology.
In response to the PCAST report, the FCC issued a Notice of Proposed Rule Making (NPRM) in 2012, which was followed by a Further Notice of Proposed Rule Making (FNPRM) \cite{FCCFNPRM} in 2014. In particular, these FCC notices proposed spectrum sharing in the 3550 - 3650 MHz band used by radars. The FNPRM included a recommendation for the establishment of exclusion zones that extend many kilometers inland from coastal areas where the radars operate. Spectrum sharing would be prohibited in these exclusion zones, as a means to prevent interference between the radars and commercial or other users. However, as we demonstrate, co-existence of wireless systems is more complex than just the transmit powers and requires detailed study and analysis through rigorous field trials to determine the accurate impact of interfering signals on commercial communications. 

In this paper, we report a first-ever experimental study on the co-existence between radar and Long Term Evolution (LTE) systems in the 3.5GHz band. The experiments presented here demonstrate, overestimation of the exclusion zones for the prevention harmful interference to commercial LTE systems, blindly based on transmit power of interference, is unwarranted. Our results suggest that detailed analysis of co-existence for two wireless systems would be more valuable than rules based on apparent system characteristics and intrinsic assumptions. \vspace{-12pt} 

\section{Experimental Setup}

To characterize the ability of Time-Division Long-Term Evolution (TD-LTE) to operate with interference from in-band radar in the 3550-3650 MHz frequency band, a series of LTE links were established near a high power radar. The experiment consisted of a TD-LTE link operating in the presence of a SPN-43C whose center frequency was set to 3571 MHz. Block error-rate (BLER) and throughput were the primary figures of merit used to quantify how well the 4G LTE system performed. Measurements were taken as a function of frequency offset from the radar's center frequency and under various operating conditions as described later in this paper. \vspace{-12pt}

\subsection{Incumbent Radar: AN/SPN-43C}

The incumbent radar was a land-based AN/SPN-43C radar at NOLF Webster Field (annex to Naval Air Station Patuxent River) in St. Inigoes Maryland. The AN/SPN-43C is a ship borne air traffic control radar that provides real time aircraft surveillance, identification, and landing assistance data. It has a nominal peak pulsed power of 1 MW and measured antenna gain of 33.4 dBi \cite{Mearns14}. SPN-43C has a range of 300 yards to 50 nautical miles and an altitude span of 30,000 ft. Typical specifications of the radar are reported in Table~\ref{tRadarspecs}\cite{CD14}.

\begin{table}
\begin{footnotesize}
\begin{center}
\caption{\small Specifications of AN/SPN-43C}
\label{tRadarspecs}
\begin{tabular}{|l|l|}
\hline
\multicolumn{2}{|l|}{\bf Transmitter}\\
\hline
Tuning Range	& 3.50-3.65 GHz\\
\hline
Pulse Generation Method	& Magnetron\\
\hline
Pulse Interval  &	889 ($\pm$20) $\mu$s\\
\hline
Pulse Width	 & 0.9 ($\pm$0.15) $\mu$s\\
\hline
Power &	850 ($\pm$150) kW\\
\hline
Bandwidth (Nominal) & 1.3 MHz\\
\hline
\multicolumn{2}{|l|}{{\bf Antenna} (Mechanical Up-Tilt = 3$^0$)}\\
\hline
Polarization &	Horizontal or circular, switchable\\
\hline
Boresight Gain	& 32 dBi\\
\hline
Rotation Period & 4s (15 rpm)\\
\hline
\multicolumn{2}{|l|}{{\bf Operation Mode:} Regular ``Search" Mode}\\ 
\hline
\end{tabular} \vspace{-18pt}
\end{center}
\end{footnotesize}
\end{table}

\vspace{-12pt}
\subsection{Measurement Sites}

\begin{figure}
\centering
\begin{subfigure}{0.5\textwidth}
  \centering
  \includegraphics[width=0.8\linewidth]{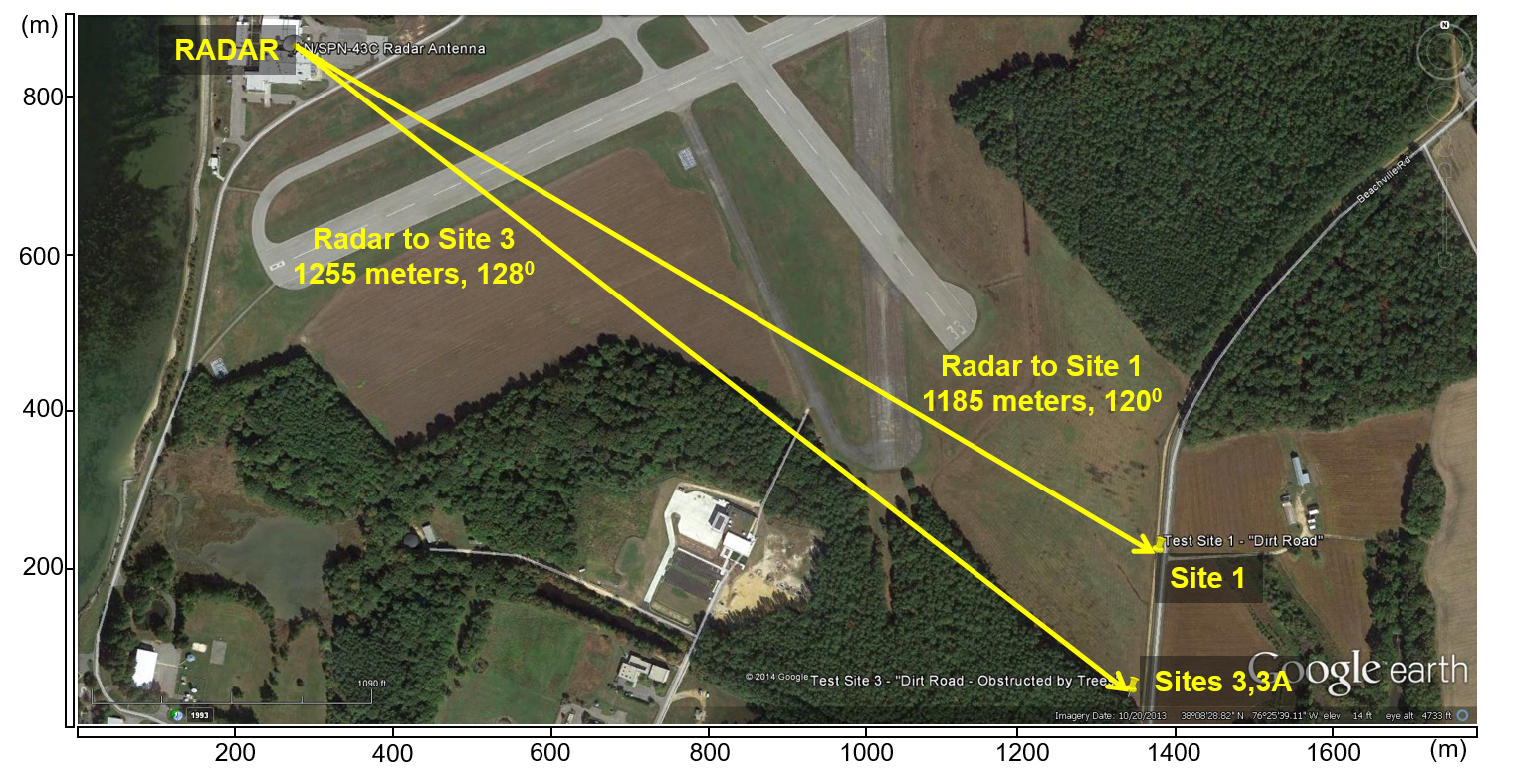}
  \caption{}
\end{subfigure}\\
\begin{subfigure}{0.5\textwidth}
  \centering
  \includegraphics[width=0.8\linewidth]{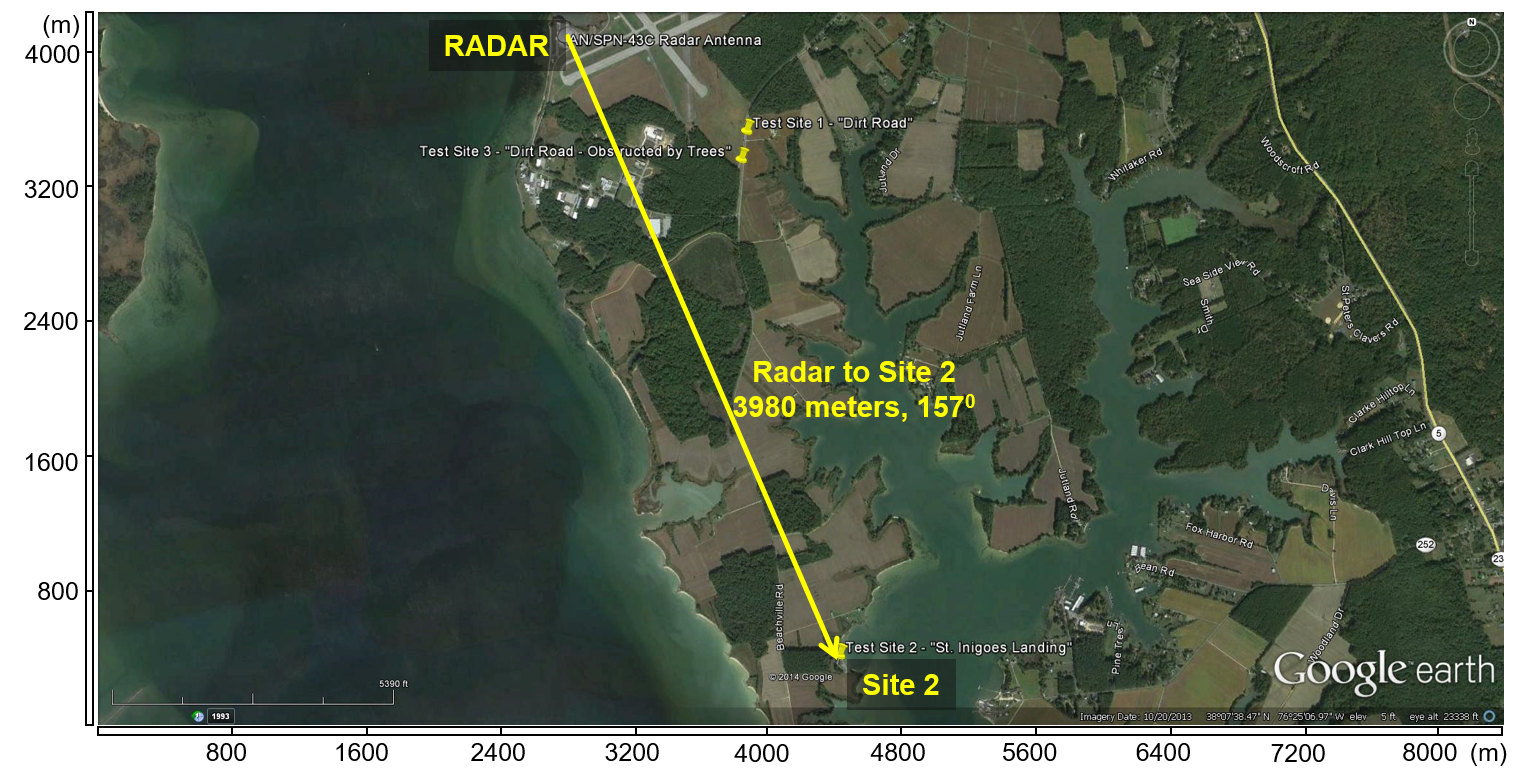}
  \caption{}
\end{subfigure}
\caption{\small Location of measurement Sites  (a) Sites 1, 3 and 3A on NOLF Webster Field and (b) Site 2 at St. Inigoes Landing \vspace{-18pt}}
\label{floc}
\end{figure}

Measurements were taken from three different locations to provide some variability in the conditions under which the LTE system was to operate. Most importantly, the received power of the interfering radar signal varies between sites. These sites were selected to include a scenario where the radar was in direct line-of-sight of the LTE system at a distance of approximately 1 km, where the radar was obstructed by forest at a similar distance, and where it was heavily obstructed by clutter at a distance of approximately 4 km. Radar power density (dBm/cm$^2$) values were measured at each of these locations. Measurements from Site 3 were repeated with the radar up-tilt lowered to the horizon, thereby increasing the power of the radar signal received by the LTE equipment. The repeated Site 3 measurements were designated Site 3A. Fig.~\ref{floc} shows the geographic relation of the radar system and the measurement sites. Since Sites 3 and 3A had a band of trees blocking the direct radar path, the radar energy was received primarily by scattering. Vertically polarized power density components are reported in Table II.

\begin{table*}
\begin{footnotesize}
\begin{center}
\caption{\small Measurement Variables}

\begin{tabular}{|c|c|c|c|c|c|c|}

\hline
Site & Distance from & Azimuth to  & Direction of eNB  & Obstruction in & Radar Antenna & 	Radar Power  \\
Location & Radar (km) & Radar (deg.) & to UE Path (deg.) &  Radar Path & Elevation (deg.)  & Density (dBm/cm$^2$) \\
\hline
1 & $1.185$ &	 $300$ & $190$ & unobstructed & $+3$ & $-17.1$\\
\hline
2 & $3.980$ & $337$ & $170$ & forest & $+3$ & $-84.6$\\
\hline
3 & $1.255$ &	$308$ & $70$ & forest & $+3$ & $-57.4$\\
\hline
3A & $1.255$ & $308$ & $70$ & forest & $0$ & $-14.8$\\
\hline

\end{tabular}\vspace{-18pt}
\end{center}
\end{footnotesize}
\end{table*}

\vspace{-12pt}
\subsection{TD-LTE Measurement Equipment}
\begin{figure}
  \centering
    \includegraphics[width=0.65\linewidth]{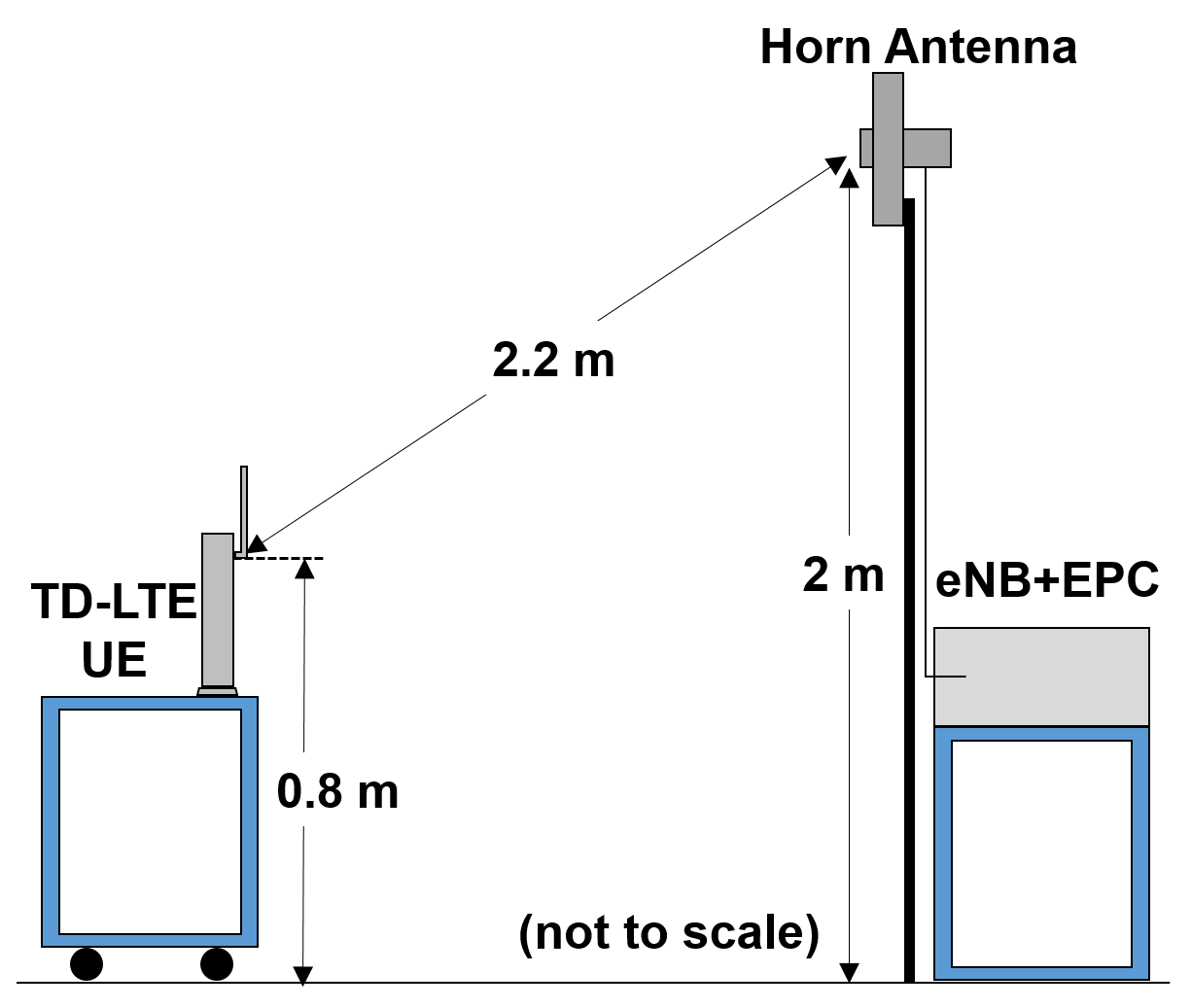}\vspace{-6pt}
  \caption{\small Diagram of the TD-LTE testbed \vspace{-24pt}}
  \label{ftdltetestbed}
\end{figure}

The 4G LTE system consisted of a single base station (eNB), a simplified evolved packet core (EPC), and a single device for user equipment (UE). Details of the specific devices are included in the below sub-sections followed by a depiction of the physical setup during the experiment in Fig.~\ref{ftdltetestbed}.

{\it Base Station:} A Rhode \& Schwarz CMW500 served as the eNB as well as a source for the measurements recorded throughout the experiment. Operating in TD-LTE test mode, the CMW500 emulates an eNB and simplified EPC. The CMW500 can connect to any device that supports the LTE standards (3GPP Release 8 and 9). Furthermore, it can provide a number of different measurements related to the link quality as well as RF metrics of the received signal. While the CMW500 has sufficient processing power and knowledge of LTE protocols and waveforms to act as an eNB, it is intended to be used with wired connections and lacks a power-amplifier or low-noise-amplifier. As such, it has much lower maximum output power and more limited receiver sensitivity than a typical eNB that is designed for over-the-air experiments. This limitation can be offset somewhat by modifying an “External Attenuation” setting which causes the CMW500 to adjust its gain stages to compensate proportionally for external losses.  Values of up to 50 dB may be entered to adjust the transmitter and receiver gains. The maximum available transmitting power of eNB was $-$15 dBm (in the full signal bandwidth).

{\it Base Station Antenna:} A Chaparral Polorotor\textsuperscript{\tiny{TM}} 1 E/A ``Scalar Horn" reflector feed was used as an antenna for the eNB. This antenna has a wide beam radiation pattern with a rapid fall off at the beam edges. The horn f/D setting was 0.42. With this setting the eNB antenna had an approximate gain of 10 dBi. The antenna was set for vertical polarization to match the antenna polarization of the user equipment (UE).  The antenna was fed through approximately 2 meters of $\frac{1}{4}$-inch Heliax\textsuperscript{\tiny{TM}} cable and a short pigtail of LDF-100 cable.  The antenna was mounted on a 2 meter tall tripod mast.
	
\looseness-1 {{\it User Equipment:} A Huawei B593s-42 wireless router supporting Band 42 (3400-3600 MHz) was selected as the TD-LTE user equipment (UE) because it was the only available commercial product whose operating band overlapped radar's center frequency. Initially, power limiters were installed on both main and diversity antenna ports to protect the receiver from potential damage from the high power radar signal. Measurement results indicated that the power limiter on the main antenna reduced the transmitted power of the UE to $-$8 dBm. Only the final measurement session (at site 3A) was performed without the power limiters in place, which allowed a transmit power of 20 dBm from the UE. The vertically polarized UE antennas were assumed to have a gain of 0 dBi.}
	
{\it Overall TD-LTE Testbed Setup:} The test setup is illustrated in Figure 2. The UE was placed on a metal cart at a height of approximately 80 cm in front of the eNB antenna. The slant path (eNB antenna to UE antenna) was set to approximately 2.2 meters in each test case. This sufficed since we were primarily interested in testing whether LTE could withstand the pulse-based interference from a radar regardless of link distance. Moreover, this scenario can be easily extrapolated for femto-cell access points, which are expected to dominate in this band. A directional antenna was used on the eNB and two omni-directional antennas were provided with the commercial UE. The main UE antenna operated for both transmit and receive. The second UE antenna (not shown) operated only for reception, i.e. diversity antenna. The radar direction was nearly perpendicular to the direction of the test link path so that the eNB horn antenna provided some discrimination against the radar interference on the uplink.

\subsection{Communication Link Details}

Table III summarizes the communication link details. For this experiment the eNB and UE operated in TD-LTE mode using 10-MHz bandwidth on both the uplink and downlink. As listed in Table III, the eNB downlink modulation scheme was 16 QAM with fifty active resource blocks while the uplink modulation scheme was QPSK with fifty active resource blocks for Site 1 but only 10 resource blocks for the remaining sites. The maximum throughput corresponding to these settings is 6.408 Mbit/s on the downlink and 6.408 or 0.349 Mbit/s on the uplink. The transmit power of the downlink was $-$15 dBm and either $-$8 dBm or +20 dBm (see Table III) on the uplink (full bandwidth power levels). Path loss was directly measured between the eNB and the UE (including antenna gains as part of the path) with all antennas maintained in their nominal locations and orientations. The direct measurement showed a loss of 40.5 dB. This value compared well to the calculated free-space path loss from the Friis equation of 40.2 dB with the assumed antenna gain of 10 dB at the eNB and 0 dBi at the UE. The UE and eNB have different receiver noise floors and are consistent with the units' specifications.

\begin{table}
\begin{footnotesize}
\begin{center}
\caption{\small TD-LTE Communication Link Details}

\begin{tabular}{|c|c|c|c|c|}

\hline
	& Downlink & Uplink & Uplink & Uplink \\
	& (all sites) & (site 1) & (sites 2 and 3) & (site 3A)\\
\hline
Modulation & 16 QAM & QPSK& QPSK & QPSK\\
\hline
LTE Cell & \multirow{2}{*}{$10$} & \multirow{2}{*}{$10$} & \multirow{2}{*}{$10$} & \multirow{2}{*}{$10$} \\
Bandwidth (MHz)  & & & &\\
\hline
\# Resource Blocks& $50$ & $50$ & $10$ & $10$\\
\hline
Tx. Power per& \multirow{3}{*}{$-42.8$} & \multirow{3}{*}{$-35.8$} & \multirow{3}{*}{$-29$} & \multirow{3}{*}{$-1$} \\
Resource Element  & & & &\\
(dBm/$15$ kHz) & & & &\\
\hline
Total Tx. Power& \multirow{3}{*}{$-15$} & \multirow{3}{*}{$-8$} & \multirow{3}{*}{$-8$} & \multirow{3}{*}{$20$} \\
in Bandwidth & & & & \\
(dBm)& & & & \\
\hline
Receiver Noise  & \multirow{3}{*}{$-126$} & \multirow{3}{*}{$-81$} & \multirow{3}{*}{$-81$} & \multirow{3}{*}{$-81$}\\
Floor in $15$ kHz & &  & & \\
BW (dBm) & & & & \\
\hline
Path Loss (dB) & $40.5$ & $40.5$ & $40.5$ & $40.5$ \\
\hline
\end{tabular}\vspace{-18pt}
\end{center}
\end{footnotesize}
\end{table}

LTE was tested at 23 different frequencies centered at 3571 MHz (radar's center frequency) with $\pm$ 11 1-MHz offset increments. Uplink and downlink block error rate (BLER) and throughput were recorded for each frequency, averaged over several seconds. 

{\looseness-1 Table IV shows calculated signal to noise ratios (SNR) and signal to interference ratios (SIR) for each location. In all cases SNR was maintained at a high level. SIR was very low for all cases except Site 2 where the signal was stronger than the radar interference on both uplink and downlink. SNR was measured using a 15 kHz noise bandwidth. SIR was taken as the ratio of peak received radar power to received signal power in the full 10 MHz bandwidth. SIR was an instantaneous quantity rather than an average taken over some period of time. Negative signal to interference SIR values indicate that the interference is stronger than the desired signal when in-band. Interference power was computed from radar power density found for each site (see Table II). Since the downlink signal power and signal path were also approximately constant during the experiments, only a single downlink SIR value was found for each site. The uplink SIR were calculated using free space path loss since the angle of arrival of the scattered radar energy in relation to the eNB antenna pattern was unknown.

\begin{table}
\begin{footnotesize}
\begin{center}
\caption{\small Calculated SNR and SIR}

\begin{tabular}{|c|c|c|c|c|}

\hline
\multirow{2}{*} {Site} & \multicolumn{2}{c|}{Downlink} & \multicolumn{2}{c|}{Uplink}\\
\cline{2-5}
 & SNR (dB) & SIR (dB) & SNR (dB) & SIR (dB) \\
\hline
1 & $43$ & $-50.8$ & $45$ & $-48.4$ \\
\hline
2 & $43$ & $16.7$ & $52$ & $19.1$ \\
\hline
3 & $43$ & $-10.5$ & $52$ & $-8.1$ \\
\hline
3A & $43$ & $-38.0$ & $80$ & $-7.5$\\
\hline
\end{tabular}\vspace{-18pt}
\end{center}
\end{footnotesize}
\end{table}

\vspace{-12pt}
\section{Experimental Results}

Plots of measured block error rate (BLER) and throughput are presented in this section with the x-axis being the LTE center frequency relative to the operating center frequency of the radar. In some cases, the LTE system was unable to establish a link at all because of the severity of the interference presented by the radar. 

An additional y-coordinate was added to the BLER plots to differentiate between these no-link cases and those in which the LTE system established a link but had a BLER of 100\%. This coordinate is above 100\% and is labeled ``Link not established". Since control signals must be exchanged on both uplink an downlink, interference on either will prevent a connection. Measurements were taken for both downlink and uplink at the same time so the ``link not established" points usually coincided for both links. Where repeated measurements were taken at a given offset frequency, the resulting values were averaged for the plot. Note that the curves are not symmetric around the center frequency because of the spectral asymmetry of radar pulse and the out of band emissions (which are much stronger than LTE signals, albeit out of band). \vspace{-12pt}

\subsection{Measured Downlink Performance}

Fig.~\ref{fDLBERvsFreqOff} shows the downlink BLER for the four measurement sites. At Site 2, where the downlink SIR was +17 dB, zero errors were measured at any offset frequency. At each of the other sites there were either significant errors or a connection failure when the radar and LTE signals fully or partially overlapped.  Where there was no overlap the error rate dropped rapidly to a low value. The corresponding downlink throughput for the same four sites is shown in Fig.~\ref{fTputvsFreqOff} Throughput largely mirrors the BLER. Site 2 at 4 km distance from the radar shows a high downlink throughput at each offset frequency.  Sites 1, 3 and 3A each either failed to connect or have a significant degradation in throughput where the radar overlaps with the LTE signal band. Where the radar fell outside of the LTE band the effect of the radar was usually small or zero.  The throughput data point at 9 MHz offset was an outlier. 

\begin{figure}
\centering
\includegraphics[width=0.9\linewidth]{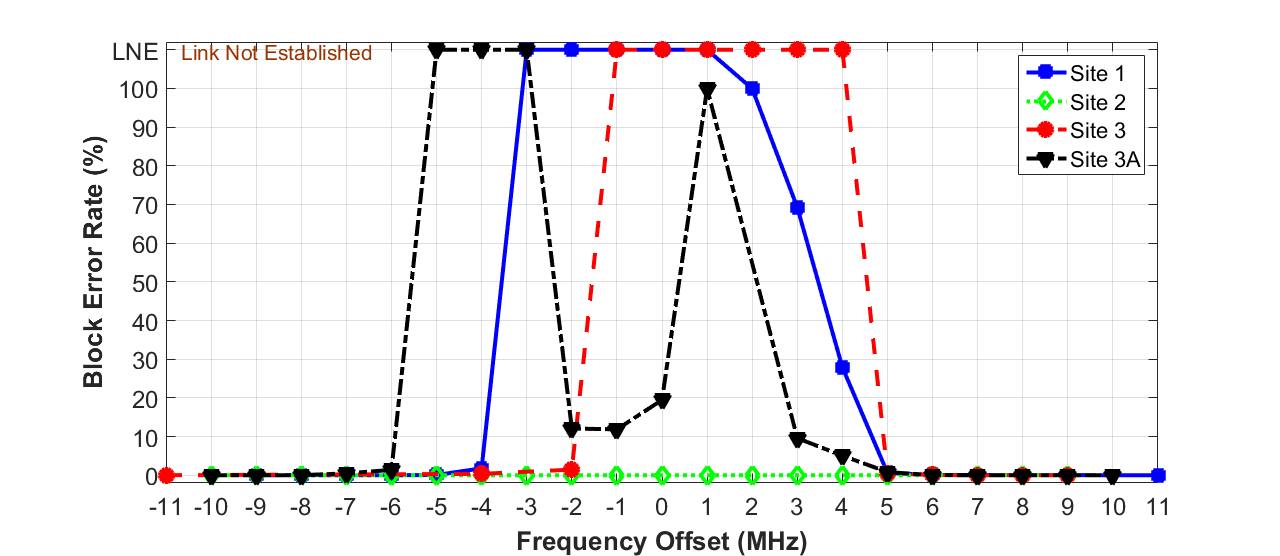}\vspace{-6pt}
\caption{\small Downlink BLER vs. Frequency offset \vspace{-12pt}}
\label{fDLBERvsFreqOff}
\end{figure}

\begin{figure}
\centering
\includegraphics[width=0.9\linewidth]{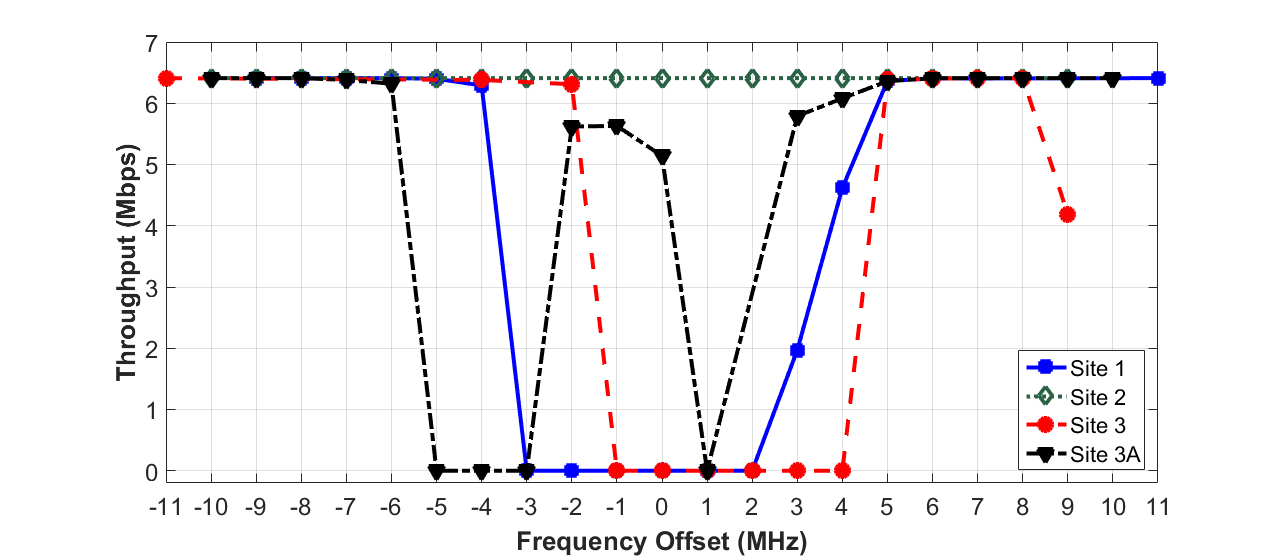}\vspace{-6pt}
\caption{\small Downlink Throughput vs. Frequency offset \vspace{-18pt}}
\label{fTputvsFreqOff}
\end{figure}

The relatively high throughput at small offsets at 3A may have been, in part, due to the higher uplink power at that site making the initial connection possible. Even with direct line of sight to the radar at a distance of about 1 km (Site 1), it was possible to maintain high data rates on the downlink at frequency offsets of 6 MHz or greater from the radar. \vspace{-12pt}

\subsection{Measured Uplink Performance}

Fig.~\ref{fULBERvsFreqOff} shows how uplink BLER varies as a function of frequency offset. The x-axis is the LTE center frequency relative to the radar center frequency (approximately 3571 MHz). Similar to the downlink cases, where the LTE system was unable to connect a point is plotted on the line above 100 percent labeled ``Link not established".  Site 2, at a distance of about 4 km from the radar, had a nearly zero uplink BLER at any frequency. No evidence of the radar interference was noted at that location. At other locations the results were similar to the downlink cases. While it was sometimes possible to transmit uplink data when the radar was located within the LTE frequency band (even at 1 km distances), BLER dropped rapidly to near zero values when the radar was outside of the LTE uplink passband.

\begin{figure}
\centering \vspace{-12pt}
\includegraphics[width=0.9\linewidth]{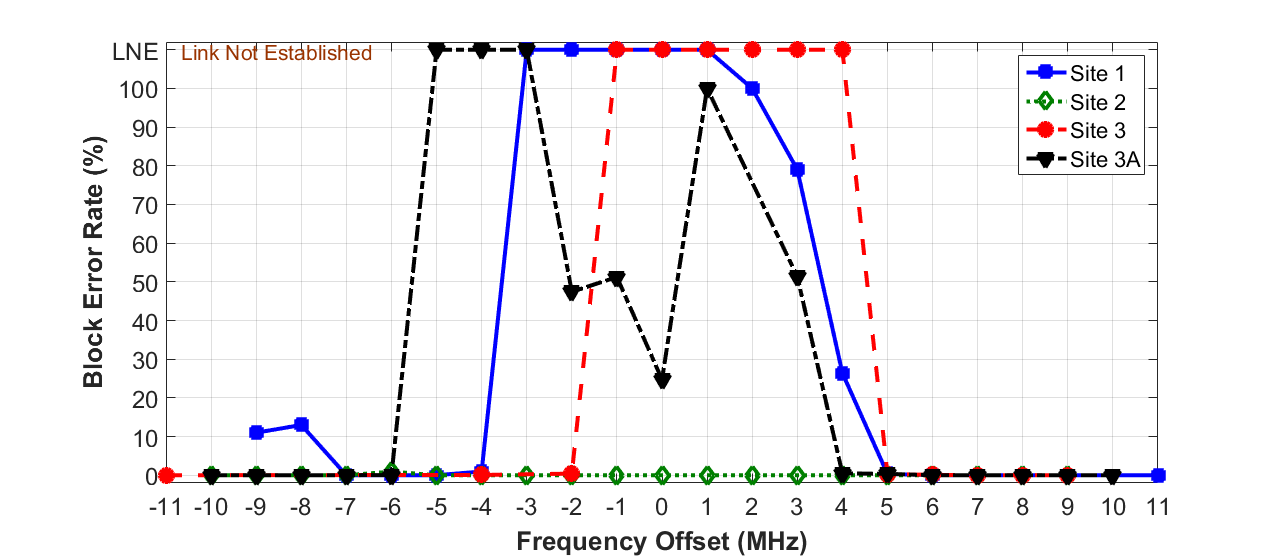}
\caption{\small Uplink BLER vs. Frequency offset}%
\label{fULBERvsFreqOff}
\end{figure}

Figures \ref{fULTputvsFreqOffa} and \ref{fULTputvsFreqOffb} show uplink throughput. Site 1 was put on a different plot since it had a different maximum throughput because there were 50 active resource blocks rather than the 10 that were used at the other sites. In all cases except Site 2 there was either a reduction of throughput or a loss of connection where the radar overlapped the LTE band. The apparent asymmetry of the throughput at frequency points near the center of the plot is probably due to the use of only 10 resource blocks grouped near the lower end of the LTE uplink transmission band. Although the radar signal was within the nominal 10 MHz LTE uplink band, it did not necessarily fall on the resource elements that were actually in use.

\begin{figure}
\centering
\includegraphics[width=0.9\linewidth]{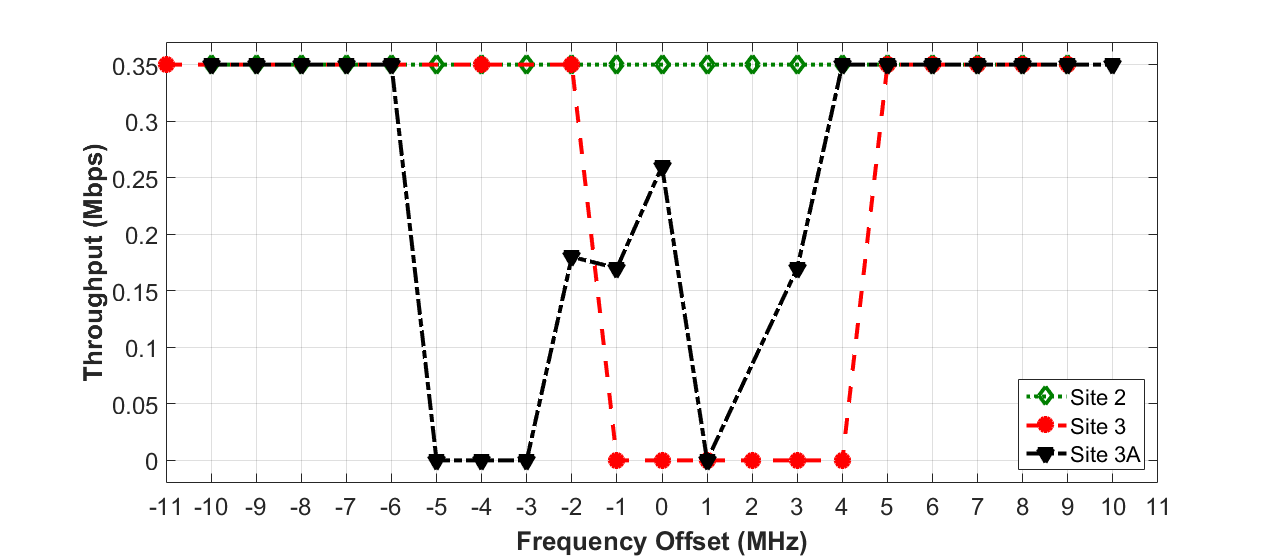}\vspace{-6pt}
\caption{\small Uplink Throughput vs. Frequency offset: Sites 2, 3 and 3A.\vspace{-12pt}}
\label{fULTputvsFreqOffa}
\end{figure}

\begin{figure}
\centering
\includegraphics[width=0.9\linewidth]{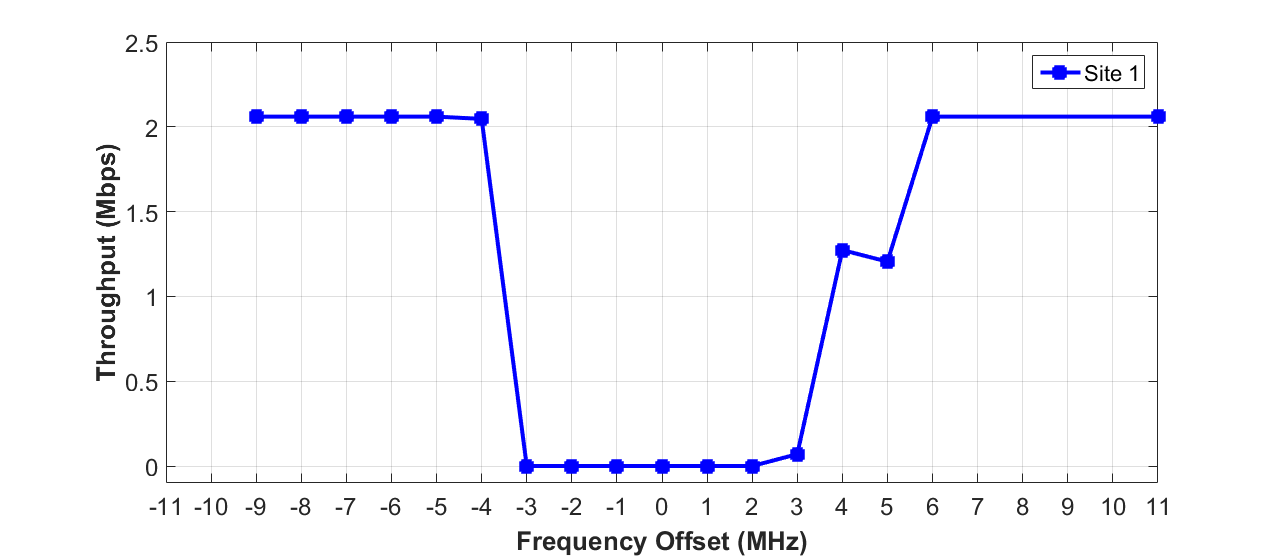}\vspace{-6pt}
\caption{\small Uplink Throughput vs. Frequency offset: Site 1. \vspace{-12pt}}
\label{fULTputvsFreqOffb}
\end{figure}

\vspace{-6pt}
\section{Conclusions}

The results of the field experiment were favorable for spectrum sharing, especially considering the close proximity of the test sites to the radar. The results specific to this experiment indicate that dynamic-spectrum-access-enabled LTE systems can avoid using overlapping frequencies with the radar and operate adjacent to those frequencies with a small guard band at close distances of 1-3 km. We found that beyond 4-5 km, the communications link operated with zero measured errors even when the carrier was directly centered on the radar frequency. In general, for co-existence analysis detailed study of the signal types, radio technologies, terrain, and propagation among other parameters is imperative before any conclusions are drawn. From the perspective of the LTE system, only a very small exclusion zone is required so long as overlap with the actual radar frequency is avoided. Exclusion zones should to be drawn after careful considerations of the compatibility of the wireless systems. Future work includes possible improvements on LTE performance with adaptive scheduling and allocation for overlapping frequencies. 

\end{document}